\documentclass[%
  twocolumn,
 showpacs,
 showkeys,
 preprintnumbers,
 amsmath,amssymb,
 aps,
  pra,
  longbibliography,
 ]{revtex4-1}

\usepackage{tikz}
\usepackage[breaklinks=true,colorlinks=true,anchorcolor=blue,citecolor=blue,filecolor=blue,menucolor=blue,pagecolor=blue,urlcolor=blue,linkcolor=blue]{hyperref}
\usepackage{graphicx}% Include figure files
\usepackage{url}

\usepackage{xcolor}

\usepackage{eurosym}

\begin{document}

\title{Reflexive measurements, self-inspection and self-representation}

%\cdmtcsauthor{Karl Svozil}
%\cdmtcsaffiliation{Vienna University of Technology}
%\cdmtcstrnumber{407}
%\cdmtcsdate{September 2011}
%\coverpage

\author{Karl Svozil}
\affiliation{Institute for Theoretical Physics, Vienna
    University of Technology, Wiedner Hauptstra\ss e 8-10/136, A-1040
    Vienna, Austria}
\affiliation{Department of Computer Science, University of Auckland, Private Bag 92019,  Auckland 1142, New Zealand}
\email{svozil@tuwien.ac.at} \homepage[]{http://tph.tuwien.ac.at/~svozil}

%\preprint{CDMTCS preprint nr. 407/2011}

\begin{abstract}
There exist limits of self-inspection due to self-referential paradoxes, incompleteness and fixed point theorems. As quantum mechanics dictates the exchange of discrete quanta, measurements and self-inspection of quantized systems are fundamentally limited.
\end{abstract}

\maketitle

\section{Embedded observers and self-expression}

The study of physics and science in general~\cite{sosa-rk1,sosa-rk2} and, in particular,
any attempt to measure a physical state while at the same time not disturbing this very state,
resembles the Baron Munchausen pulling himself and the horse on which he was sitting out of a mire by his own hair.
Physics is grounded in reflexivity and self-inspection and is bound to epistemology; there is no direct access to ontology
(any claim regarding the latter is metaphysical and remains conjectural).
Empirical evidence can solely be drawn from operational
procedures accessible to embedded~\cite{toffoli:79,svozil-94,svozil-93,roessler-87,roessler-98} observers.
Embeddedness means that intrinsic observers have to somehow interact with the object,
thereby altering both the observer as well as the object inspected.

Physics shares this feature with computer science as well as the formalist,
axiomatic approach to mathematics; actually, this article could be
interpreted as a prolegomenon to formal (in)completeness,
as subsumed by Lawvere~\cite{Lawvere1969} and Yanofsky~\cite{Yanofsky-BSL:9051621}.
There, consistency requirements result in limits of self-expressivity
relative to the axioms (if the formal expressive capacities are ``great enough'').
Indeed, as expressed by
G\"odel (cf. Ref.~\cite[p.~55]{v-neumann-66} and \cite[p.~554]{fef-84}),
{\em ``a complete epistemological description
 of a language $A$ cannot be given in the same language $A$, because
 the concept of truth of sentences of $A$ cannot be defined in $A$. It
 is this theorem which is the true reason for the existence of
 undecidable propositions in the formal systems containing arithmetic.''}

A generalized version of Cantor's theorem suggests that non-trivial
(that is, non-degenerate, with more than one property)
systems cannot intrinsically express all of its properties.
For the sake of a formal example~\cite[p.~363]{Yanofsky-BSL:9051621},
take any set $\textsf{\textbf{S}}$ and some (non-trivial, non-degenerate) ``properties'' $\textsf{\textbf{P}}$ of $\textsf{\textbf{S}}$.
Then there is no onto function $\textsf{\textbf{S}} \longrightarrow \textsf{\textbf{P}}^\textsf{\textbf{S}}$,
or equivalently~\footnote{
Every function
$f: \textsf{\textbf{S}} \longrightarrow \textsf{\textbf{P}}^\textsf{\textbf{S}}$
can be converted into an equivalent function $g$,
with
$g: \textsf{\textbf{S}} \times \textsf{\textbf{S}} \longrightarrow \textsf{\textbf{P}}$,
such that $g(a_1,a_2) = [f (a_2)](a_1) \in \textsf{\textbf{P}}$.
One may think of
$a_2$ as some ``index'' running over all functions $f$.

A typical example is taken from Cantor's proof that the (binary) reals are non-denumerable:
Identify $\textsf{\textbf{S}}= {\Bbb N}$ and $\textsf{\textbf{P}}=\{0,1\}$, then $\{0,1\}^{\Bbb N}$ can be identified with the
binary reals in the interval $[0,1]$.
Any function $f(n) =  r_n$ with $n \in {\Bbb N}$ and $r_n \le \in [0,1]$
representable in index notation as $r_n=0.r_{n,1}r_{n,2}\cdots r_{n,k} \cdots $
can be rewritten as $[f(n)](k) = g(n,k) = r_{n,k}$.
}
$\textsf{\textbf{S}} \times \textsf{\textbf{S}} \longrightarrow \textsf{\textbf{P}}$, whereby
$\textsf{\textbf{P}}^\textsf{\textbf{S}}$ represents the set of functions from $\textsf{\textbf{S}}$ to $\textsf{\textbf{P}}$.
Stated differently, suppose some (nontrivial, non-degenerate) properties; then
the set of all conceivable and possible functional images or ``expressions'' of those properties
is strictly greater than the domain or ``description'' thereof.

For the sake of construction of a ``non-expressible description'' relative to the set of
all functions $f: \textsf{\textbf{S}} \longrightarrow \textsf{\textbf{P}}^\textsf{\textbf{S}}$,
let us closely follow Yanovsky's scheme~\cite{Yanofsky-BSL:9051621}:
suppose that, for some non-trivial set of properties $\textsf{\textbf{P}}$ we can define (that is, there exists)
a ``diagonal-switch'' function
$\delta: \textsf{\textbf{P}} \longrightarrow \textsf{\textbf{P}}$
without a fixed point,
such that, for all $p\in \textsf{\textbf{P}}$, $\delta (p)\neq p$.
Then we may construct a non-$f$-expressible function $u: \textsf{\textbf{S}} \longrightarrow \textsf{\textbf{P}}^\textsf{\textbf{S}}$ by forming
\begin{equation}
u(a) = \delta(g(a,a)),
\end{equation}
with $g(a,a) = [f(a)](a)$.

Because, in a proof by contradiction, suppose that some function $h$ expresses $u$;
that is,  $u(a_1) = h(a_1,a_2)$.
But then, by identifying $a=a_1=a_2$, we would obtain
$h(a,a)=\delta(h(a,a))$,
thereby contradicting our definition of $\delta$.
In summary, there is a limit to self-expressibility as long as one deals with
systems of sufficiently rich expressibility.

\section{Reflexive measurement}

In a very similar manner we identify $A$ with measurements $\textsf{\textbf{M}}$,
and $P$ with the set of possible outcomes $\textsf{\textbf{O}}$ of these measurements.
Alternatively, we may associate a physical state with $P$.

For the sake of construction of a ``non-measurable self-inspection'' relative to
all operational capacities
let us again closely follow the scheme involving the non-existence of fixed points.
In particular, let us assume that it is not possible to measure properties without changing them.
This can be formalized by
introducing a {\em disturbance function}
$\delta : \textsf{\textbf{O}} \longrightarrow \textsf{\textbf{O}}$
without a fixed point,
such that, for all $o\in \textsf{\textbf{O}}$, $\delta (o)\neq o$.
Then we may construct a non-operational measurement
$u: \textsf{\textbf{M}} \longrightarrow \textsf{\textbf{O}}^\textsf{\textbf{M}}$
by forming
\begin{equation}
u(m) = \delta(g(m,m)),
\end{equation}
with $g(m,m) = [f(m)](m)$.

Again, because, in a proof by contradiction, suppose that some operational measurement $h$ could express $u$;
that is,  $u(m_1) = h(m_1,m_2)$.
But then, by identifying $m=m_1=m_2$, we would obtain
$h(m,m)=\delta(h(m,m))$,
thereby again clearly contradicting our definition of $\delta$.

In summary, there is a limit to self-inspection, as long as one deals with
systems of sufficiently rich phenomenology.
One of the assumptions has been that there is no empirical self-exploration and self-examination without
changing the sub-system to be measured. Because in order to measure a subsystem, one has to interact with it;
thereby destroying at least partly its original state.
This has been formalized by the introduction of a ``diagonal-switch'' function
$\delta: \textsf{\textbf{P}} \longrightarrow \textsf{\textbf{P}}$
without a fixed point.

In classical physics one could argue that, at least in principle, it would be possible
to push this kind of disturbance to arbitrary low levels,
thereby effectively and for all practical purposes (FAPP)
eliminating the constraints on and limits from self-observation.
One way of modelling this would be a double pendulum; that is,  two coupled oscillators,
one of them (the subsystem associated with the ``observed object'') with a ``very large'' mass,
and the other one of them (the subsystem associated with the ``observer'' or the ``measurement apparatus'')
with a ``very small'' mass.

In quantum mechanics this possibility is blocked by the discreteness of the exchange of at least one single quantum of action.
Thus there is an insurmountable quantum limit to the resolution of measurements,
originating in self-inspection.

\section{Intrinsic self-representation}

Having now explored the limits and the ``negative'' effects of the type of self-exploration and self-examination
embedded observers are bound to we shall now examine the ``positive'' side of self-description.
In particular, we shall proof that, at least for ``nontrivial''
deterministic systems (in the sense of recursion theory and,
by the Church-Turing thesis, capable of universal computation),
it is possible to represent a complete theory of itself within this very system.

To avoid any confusion one must differentiate between determinism and predictability.
As has already been pointed out by Suppes~\cite{suppes-1993}, any embodiment of a Turing machine,
such as in ballistic $n$-body computation~\cite{svozil-2007-cestial}
is deterministic; and yet, due to the recursive undecidability of the halting problem,
certain aspects of its behaviour, or phenomenology, are unpredictable.

The possibility of a complete formal representation of a non-trivial system (capable of universal computation)
within that very system is a consequence of the recursion theorem~\cite{Yanofsky-BSL:9051621}
and Kleene's s-m-n theorem:
Denote the partial function $g$ that is computed by the Turing-machine
program with description $i$ by $\varphi_i$.

Suppose that $f: {\Bbb N} \longrightarrow {\Bbb N}$ is a total (defined on its entire domain) computable function.
Then there exists an $n_0 \in {\Bbb N}$ such that $\varphi_{f(n_0)}=\varphi_{n_0}$.
For a proof, see Yanofsky~\cite{Yanofsky-BSL:9051621}.

The s-m-n theorem states that every partial recursive function $\varphi_i (x,y)$
can be represented by a total recursive function $r(i,x)$ such that
$\varphi_i (x,y)= \varphi_{r(i,x)} (y)$,
thereby hard-wiring the input argument $x$ of $\varphi_i (x,y)$ into the index of $\varphi_{r(i,x)}$.

Now we are ready to state that
a complete formal representation or description of a non-trivial system (capable of universal computation)
is given by the number ${n_0}$ of the computable function $\varphi_{n_0}$ which always (that is,
for all input $x$) outputs its own description; that is, $\varphi_{n_0} (x) = {n_0}$.

For the sake of a proof, suppose
that $p: {\Bbb N} \times {\Bbb N}  \longrightarrow {\Bbb N}$ is the projection function
$p(m,n)= m$.
By the s-m-n theorem there exists a totally computable function $r$ such that
$\varphi_{r(y)}(x) = p(y,x)=y$.
And by the recursion theorem,
there exists a complete description
${n_0}$ such that $\varphi_{n_0}(x)=\varphi_{r({n_0})}= p({n_0},x)= {n_0}$.

\section{Summary and relation to complementarity and the randomness postulate}

It has been argued that it is possible to have a complete theory
of any deterministic physical system which has the capacity
to embody universal computation.
How to obtain such a complete theory is an altogether different issue,
as the rule inference problem is undecidable~\cite{go-67,blum75blum,angluin:83,ad-91}.

Such complete theory or self-representation of everything is no entitlement to omniscience,
as, due to the reduction of the halting problem,
it is in general impossible to predict certain behaviours.

Moreover, self-inspection is limited by its paradoxical character,
as in general measurements  change the physical state.
This is, in particular, relevant in quantum mechanics, when discrete quanta are exchanged.
Quantum and automata complementarity~\cite{e-f-moore},
and complementarity in general~\cite{svozil-2001-eua},
are related to {\em finite} (reversible~\cite{svozil-2004-kyoto}) systems.

\acknowledgments{
This work was supported in part by the European Union, Research Executive Agency (REA),
Marie Curie FP7-PEOPLE-2010-IRSES-269151-RANPHYS grant.

%Responsibility for the information and views expressed in this article lies entirely with the author.
%The content therein does not reflect the official opinion of the Vienna University of Technology or the University of Auckland.

The author declares no conflict of interest.

}

%\bibliography{svozil}

%merlin.mbs apsrev4-1.bst 2010-07-25 4.21a (PWD, AO, DPC) hacked
%Control: key (0)
%Control: author (0) dotless jnrlst
%Control: editor formatted (1) identically to author
%Control: production of article title (0) allowed
%Control: page (1) range
%Control: year (0) verbatim
%Control: production of eprint (0) enabled
%

\end{document}